\pdfoutput=1
\documentclass[a4paper,11pt]{article}
\usepackage{amsmath}
\usepackage{amssymb}
\usepackage{graphicx}
\usepackage{color}
\usepackage[utf8]{inputenc} 

\usepackage[square,numbers,comma,sort&compress]{natbib}
\usepackage[colorlinks=true,citecolor=blue,linkcolor=purple,urlcolor=purple]{hyperref}
\usepackage{geometry}
\usepackage{verbatim}
\usepackage[usenames,dvipsnames,svgnames,table]{xcolor}
\usepackage{booktabs}
\usepackage{slashed}

\usepackage{tikz-feynman}
\tikzfeynmanset{
every edge={thick},
}
\usepackage{subcaption}
\usepackage{multirow}

\def\varabstract{ }
\def\varkeywords{ }
\def\vararxivnumber{ }
\def\vartitle{ }
\def\varsubtitle{ }
\renewcommand{\title}[1]{\gdef\vartitle{#1}}
\newcommand{\subtitle}[1]{\gdef\varsubtitle{#1}}
\renewcommand{\abstract}[1]{\gdef\varabstract{#1}}
\newcommand{\keywords}[1]{\gdef\varkeywords{#1}}
\newcommand{\arxivnumber}[1]{\gdef\vararxivnumber{#1}}
\newtoks\authtoks
\renewcommand{\author}[2][]{%
	\authtoks=\expandafter{\the\authtoks#2$^{#1}$\ }%
}
\newtoks\affiltoks
\newcommand{\affiliation}[2][]{%
    \affiltoks=\expandafter{\the\affiltoks{\item[$^{#1}$]#2}}%
}
\newtoks\emailtoks\newcounter{emailcounter}%
\newcommand{\emailAdd}[1]{%
\ifnum\theemailcounter>0\emailtoks=\expandafter{\the\emailtoks, \typeemail{#1}}%
\else\emailtoks=\expandafter{\typeemail{#1}}%
\fi
\stepcounter{emailcounter}}
\newcommand{\typeemail}[1]{\href{mailto:#1}{\tt #1}}

\renewcommand\maketitle{
	\newgeometry{margin=2cm}
	\pagestyle{empty}\setcounter{page}{0}
	{\huge\flushleft\sffamily\bfseries\vartitle\\\Large\varsubtitle\par}
\vskip6ex
{\large\bfseries\raggedright\sffamily\the\authtoks\par}
\vskip2ex
\begin{list}{}{%
\setlength{\leftmargin}{0.28cm}%
\setlength{\labelsep}{0pt}%
\setlength{\itemsep}{-3pt}%
\setlength{\topsep}{-\parskip}}
\itshape\small%
\the\affiltoks
\end{list}
\vskip2ex
\noindent\hspace{0.28cm}\begin{minipage}[l]{.9\textwidth}
\begin{flushleft}
\textit{E-mail:} \the\emailtoks
\end{flushleft}
\end{minipage}
\vskip5ex
\noindent{\renewcommand\baselinestretch{.9}\textsc{Abstract:}}\ \varabstract
\vskip5ex 
\if!\varkeywords!\else\noindent{\textsc{Keywords:}}\ \varkeywords \vskip2ex\fi
\if!\vararxivnumber!\else\noindent{\textsc{ArXiv ePrint:}} \href{http://arxiv.org/abs/\vararxivnumber}{\vararxivnumber}\vskip2ex\fi

\newpage
\restoregeometry
\pagestyle{plain}

\setcounter{footnote}{0}
} 

\usepackage[square,numbers,comma,sort&compress]{natbib}

\definecolor{MS}{rgb}{0,0,1}
\newcommand{\braket}[1]{\left\langle#1\right\rangle}

\usepackage{ifthen}
	
	\newcommand{\barlimb}[5]{
  \pgfmathparse{\mypos+0.3}
  \edef\mypos{\pgfmathresult}
		\node[left,scale=0.6] at (0,\mypos) {#1};
		\fill[#2] ($(0,\mypos)+(0,-0.1)$) rectangle +(#3,0.2);
		\node[left,scale=0.6] at (#3,\mypos) {#3};
		\fill[#4] ($(0,\mypos)+(0,-0.1)$) rectangle +(#5,0.2);
		\node[left,scale=0.6] at (#5,\mypos) {#5};
}
	\newcommand{\barlimc}[7]{
  \pgfmathparse{\mypos+0.3}
  \edef\mypos{\pgfmathresult}
		\node[left,scale=0.6] at (0,\mypos) {#1};
		\pgfmathparse{#3 > 5 ? 1 : 0}
		\ifthenelse{\pgfmathresult=1}{
			\fill[#2] ($(0,\mypos)+(0,-0.1)$) rectangle +(5,0.2);
			\fill[white] ($(0,\mypos)+(3.5,-0.1)$) rectangle +(0.3,0.2);
			\draw[decoration={zigzag},decorate,#2,very thick] (3.4,\mypos) to +(0.5,0);
			\node[left,scale=0.6] at (5,\mypos) {#3};
			}{
			\fill[#2] ($(0,\mypos)+(0,-0.1)$) rectangle +(#3,0.2);
			\node[left,scale=0.6] at (#3,\mypos) {#3};
		}		
		\fill[#4] ($(0,\mypos)+(0,-0.1)$) rectangle +(#5,0.2);
		\node[left,scale=0.6] at (#5,\mypos) {#5};
		\fill[#6] ($(0,\mypos)+(0,-0.1)$) rectangle +(#7,0.2);
		\pgfmathparse{#7 <0.3 ? 1 : 0}
		\ifthenelse{\pgfmathresult=1}{
			\node[right,scale=0.6] at (0,\mypos) {#7};
		}{
		\node[left,scale=0.6] at (#7,\mypos) {#7};
	}
}

\definecolor{LHC}{HTML}{3c6fff}
\definecolor{LHCunit}{HTML}{3cb8ff}
\definecolor{LE}{HTML}{75cd3c}
\definecolor{legend}{HTML}{fdd017}

\title{Lepton-flavour-violating gluonic operators:}
\subtitle{constraints from the LHC and low energy experiments}

\author[a]{Yi Cai,} \emailAdd{caiy36@mail.sysu.edu.cn}
\author[b]{Michael A.~Schmidt,} \emailAdd{michael.schmidt@sydney.edu.au}
\author[c]{and German Valencia}\emailAdd{german.valencia@monash.edu}

\affiliation[a]{School of Physics, Sun Yat-sen University, Guangzhou, 510275, China}
\affiliation[b]{ARC Centre of Excellence for Particle Physics at the Terascale, School of Physics, The University of Sydney, Physics Road, New South Wales 2006, Australia}
\affiliation[c]{School of Physics and Astronomy,
Monash University, Wellington Road,
Melbourne, Victoria 3800, Australia}

\abstract{Effective operators provide a model-independent
description of physics beyond the standard model that is particularly useful given the absence of any
signs of new physics at the Large Hadron Collider
(LHC). We recast previous LHC analyses to set limits on
lepton-flavour-violating gluonic effective operators of
dimension 8 and compare our results to existing limits from low-energy precision experiments. Current LHC data constrains the scale $\Lambda$ of the effective operators to be larger than $\Lambda \gtrsim 0.5 - 1.6$
TeV depending on the flavour and thus provides the most
stringent limit for all operators apart from parity-conserving operators of the form $G G\, \bar \mu P_{L,R} e$, where $\mu$-$e$ conversion in nuclei poses the most stringent constraint.
}

\arxivnumber{1802.09822}

\begin{document}
\maketitle


\section{Introduction}
The Large Hadron Collider (LHC) successfully discovered the 125 GeV
Higgs boson in 2012~\cite{ATLAS:2012gk,Chatrchyan:2012xdj} and thus confirmed the last missing piece of the
Standard Model (SM). However there are several hints for physics beyond
the SM. In particular the observation of neutrino oscillations~\cite{Fukuda:1998mi} (and
consequently the existence of massive neutrinos) showed that lepton
flavour is not conserved and lepton-flavour-violating (LFV) processes
exist. In particular charged leptons may change flavour in processes such as radiative muon decay $\mu\to e\gamma$. In  minimal models of
neutrino mass for Dirac neutrinos, such as the SM augmented by the Weinberg operator \cite{Weinberg:1979sa} or the seesaw model~\cite{Minkowski:1977sc}, the rates for LFV processes are tiny, because they are suppressed
by the smallness of the neutrino mass. This is no longer true in more general models, and there are many 
well-motivated models such as seesaw models with electroweak triplets~\cite{Magg:1980ut,Foot:1988aq} or radiative neutrino mass models~\cite{Zee:1980ai,Zee:1985id,Babu:1988ki} (See Ref.~\cite{Cai:2017jrq} for a recent review.) in which charged LFV processes are 
important. More generally, there are many known extensions of the SM which do not conserve lepton flavour and can be tested 
by studying LFV processes at the LHC. Examples include ($R$-parity violating)
supersymmetric models~\cite{Barbier:2004ez} and $Z^\prime$
models~\cite{Langacker:2008yv}. 

Low-energy precision experiments including MEG~\cite{Adam:2013mnn},
SINDRUM~\cite{Dohmen:1993mp} and the
B-factories BaBar~\cite{Aubert:2001tu} and Belle~\cite{Abashian:2000cg} have searched for charged LFV
processes and placed severe constraints on many LFV processes. At the same time, the LHC
experiments
are currently pushing the limits for the scale of any new physics higher
and higher. Effective field theory (EFT) is the ideal framework to describe new LFV processes under these conditions. 
Existing studies have mostly focused on the leading dimension six effective operators  with two quarks and two leptons. 
Ref.~\cite{Carpentier:2010ue} derived constraints from rare decays and from collider bounds on contact interactions. Ref.~\cite{Cai:2015poa} focused on LHC constraints finding they can be competitive to those from rare processes. In particular effective
LFV operators with two leptons of different flavour and two coloured
particles, quarks and gluons, provide a clean signature with low SM background and large
production cross section at the LHC, and this has been exploited in
Ref.~\cite{Cai:2015poa} to extract competitive constraints for
operators with right-handed $\tau$-leptons. 
Ref.~\cite{Bhattacharya:2014wla} suggested to use
	$pp\to \mu^\pm\tau^\mp t\bar t$ to probe operators with top
quarks and Ref.~\cite{Arganda:2015ija} studied the channel $pp\to \mu^\pm\tau^\mp jj$ to search for heavy singlet neutrinos in the inverse seesaw model. LFV operators with gluons may also be probed in a lepton-hadron collider via the process $\ell g \to \ell^\prime g$~\cite{Takeuchi:2017btl}.

In this study we focus on LFV gluonic dimension-8 operators, with two leptons of different flavour coupled to two gluons. Although technically in the EFT 
they are  suppressed with respect to dimension-6 operators, this
suppression is compensated by the large gluon content of protons at high
energies~\cite{Potter:2012yv}. Low-energy precision constraints on these operators have been
previously studied in Ref.~\cite{Petrov:2013vka}. We derive LHC limits and compare them to the updated low-energy constraints.

The paper is structured as follows: in Sec.~\ref{sec:op} we introduce
the dimension-8 operators and fix the notation. Possible ultraviolet
(UV) completions are discussed in Sec.~\ref{sec:UV}. In
Sec.~\ref{sec:LHC} we discuss LFV signals at the LHC and recast existing
LHC analyses to obtain a limit on the scale of each operator, which 
forms the main result of our study. Sec.~\ref{sec:LE} provides a brief
summary of the relevant low-energy precision constraints. Finally we
summarise in Sec.~\ref{sec:summary} and compare the sensitivity of LHC
searches with low-energy precision experiments.


\section{Effective LFV gluonic dimension-8 operators}
\label{sec:op}
There are six independent operators for
one flavour of leptons (54 for three flavours) with two gluon field strength
tensors and two leptons~\cite{Lehman:2015coa,Henning:2015alf}. The effective Lagrangian with these operators can thus be written as follows
		\begin{align}
			\mathcal{L} & = 
			x_{ij}  \mathcal{O}^{ij}_X
			+x^\prime_{ij} \mathcal{O}^{\prime ij}_X 
			+\bar x_{ij} \bar{\mathcal{O}}^{ij}_X
			+\bar x^\prime_{ij}\bar{\mathcal{O}}^{\prime ij}_X
			+ y_{ij}\, \mathcal{O}^{ij}_Y
			+ z_{ij}\, \mathcal{O}^{ij}_Z
	\end{align}
	where the Wilson coefficients $x$, $x^\prime$, $\bar x$ and $\bar x^\prime$ are real matrices and $y$ and $z$ Hermitian matrices in flavour space and the operators are defined as
	\begin{subequations}
		\label{eq:ops}	\begin{align}
		\mathcal{O}^{ij}_X&=\alpha_s G_{\mu\nu}^a G^{a\mu\nu} \left( \bar e_{Ri} L_j \cdot \phi^* + \bar L_j \cdot \phi e_{Ri} \right)  & 
		\mathcal{O}^{\prime ij}_X &= i\,   \alpha_s G_{\mu\nu}^a\tilde G^{a\mu\nu} \left(\bar e_{Ri} L_j\cdot \phi^*- \bar L_j\cdot \phi  e_{Ri} \right)\label{eq:x}\\
		\bar{\mathcal{O}}^{ij}_X & = i\, \alpha_s G_{\mu\nu}^aG^{a\mu\nu} \left( \bar e_{Ri} L_j\cdot\phi^* -\bar L_j \cdot \phi e_{Ri} \right) & 
\bar{\mathcal{O}}^{\prime ij}_X & =  \alpha_s G_{\mu\nu}^a\tilde G^{a\mu\nu} \left( \bar e_{Ri} L_j\cdot \phi^* + \bar L_j \cdot \phi e_{Ri}\right) 
\label{eq:xbar}\\ 
\mathcal{O}^{ij}_Y & =i\, \alpha_s G_{\mu\rho}^a G_{\sigma\nu}^a \eta^{\rho\sigma} \bar L_i \gamma^\mu D^\nu L_j &
\mathcal{O}^{ij}_Z & = i\, \alpha_s G_{\mu\rho}^a G_{\sigma\nu}^a \eta^{\rho\sigma} \bar e_{Ri} \gamma^\mu D^\nu e_{Rj}\;.\label{eq:yz}
\end{align}
\end{subequations}
$G_{\mu\nu}$ denotes the gluon field strength tensor and its dual $\tilde G_{\mu\nu}=\tfrac12 \epsilon_{\mu\nu\rho\sigma}G^{\rho\sigma}$. The SM Higgs doublet is denoted as $\phi$, the left-handed lepton doublet $L$ and the right-handed charged leptons $e_R$. $D_\nu$ is the covariant derivative. All operators are normalised with the strong coupling $\alpha_s\equiv g_s^2/4\pi$. The normalisation is chosen such that the operators with Wilson coefficients $x$, $x^\prime$, $\bar x$ and $\bar x^\prime$ in the first two lines are invariant under QCD corrections at one-loop order~(See e.g. \cite{Cirigliano:2009bz}).
We only consider low-energy constraints for these operators in Sec.~\ref{sec:LE}, because operators with derivatives are further suppressed at low energy. 
The combination of gluon field strength tensors $G_{\mu\nu}^a \tilde G^{a \mu\nu}$ in the primed operators $\mathcal{O}_X^\prime$ and $\bar{\mathcal{O}}_X^\prime$ violates parity which is important for low-energy constraints. The operators with an over-bar $\bar{\mathcal{O}}_X^{(\prime)}$ violate CP. Thus imposing CP results in $\bar x_{ij}=\bar x^\prime_{ij}=0$. 

The operators in Eq.~\ref{eq:ops} are defined in the weak interaction basis. We obtain the relevant interactions after a rotation of the charged leptons $e_{L,R}$ to the mass basis $\hat e_{L,R}$ and the corresponding rotation of the neutrino weak interaction eigenstates $\nu_L$\footnote{Note that the neutrino states $\hat \nu_L$ are not mass eigenstates.}
\begin{align}
	e_L &= L_e \hat e_L &
	e_R &= R_e \hat e_R &
	\nu_L & = L_e \hat \nu_L \;.
\end{align}
For simplicity we choose $L_e=R_e=1$ without loss of generality and drop the hats on the fields. In this case the leptonic mixing entirely originates from the neutrino sector. Thus the interactions of two charged leptons with two gluons are simply given by
\begin{align}
			\mathcal{L} & = 
			\frac{\alpha_s v x_{ij}}{\sqrt{2}}\, G_{\mu\nu}^a G^{a\mu\nu} \left( \bar{{e}}_{Ri}  e_{Lj} + \bar{{e}}_{Lj} {e}_{Ri} \right)  
			+\frac{i \alpha_s v  x^\prime_{ij}}{\sqrt{2}} \,G_{\mu\nu}^a\tilde G^{a\mu\nu} \left(\bar{{e}}_{Ri}  e_{Lj}- \bar { e}_{Lj} {e}_{Ri} \right)
			\nonumber \\&
			+ \frac{i \alpha_s v {\bar x}_{ij}}{\sqrt{2}} \, G_{\mu\nu}^aG^{a\mu\nu} \left( \bar{{e}}_{Ri}  e_{Lj} -\bar{{e}}_{Lj}  e_{Ri} \right) 
			+\frac{\alpha_s v {\bar x}^\prime_{ij}}{\sqrt{2}} \, G_{\mu\nu}^a\tilde G^{a\mu\nu} \left( \bar{{e}}_{Ri}  e_{Lj} + \bar{{e}}_{Lj}  e_{Ri}\right) 
			\\\nonumber &
			+  \frac{i \alpha_s v y_{ij}}{\sqrt{2}}\, G_{\mu\rho}^a G_{\sigma\nu}^a \eta^{\rho\sigma} \bar{{e}}_{Li} \gamma^\mu D^\nu {e}_{Lj} 
			+  \frac{i \alpha_s v z_{ij}}{\sqrt{2}} \, G_{\mu\rho}^a G_{\sigma\nu}^a \eta^{\rho\sigma} \bar{ e}_{Ri} \gamma^\mu D^\nu  e_{Rj}
	\end{align}
	with $\braket{\phi}=(0\;v)^T/\sqrt{2}$ and the electroweak vacuum expectation value $v\simeq 246$ GeV.


\section{UV completions}
\label{sec:UV}
In this section we discuss possible models giving
rise to the effective operators of the previous
section. Dimensional arguments suggest, as usual,
that the largest coefficients would appear in
models that produce the operators at tree-level.
The operators in Eqs.~\eqref{eq:x} and
\eqref{eq:xbar} can be produced by the s-channel
exchange of a spin zero particle in $gg \to \ell
\ell'$, whereas those in Eqs.~\eqref{eq:yz} would
be produced by the exchange of a spin two particle
in the s-channel of the same process. The two
gluons are respectively in a spin zero or two
configuration in these processes.

\begin{figure}[tb]\centering
	\begin{subfigure}{\linewidth}\centering
		\begin{tikzpicture}
	\begin{feynman}
		\vertex[blob] (a) {};
		\vertex[below left= of a] (g1) {$g$};
		\vertex[above left= of a] (g2) {$g$};
		\vertex[dot,right=of a] (b) {};
		\vertex[above right=of b] (f1) {$\ell$};
		\vertex[below right=of b] (f2) {$\ell^\prime$};
	
			\diagram {
				(g1) -- [gluon] (a) -- [gluon] (g2),
				(a) --[scalar] (b),
				(f1) [particle=\(\ell\)] -- [fermion] (b)  -- [anti fermion] (f2) [particle=\(\ell^\prime\)],
	};
\end{feynman}
\node[below] at ($(a)!0.5!(b)$) {$\frac{1}{m_H^2}$};
\node[left,blue,xshift=-5ex] (at) at (a) {$\frac{\alpha_s}{12\pi v}G^{A\mu\nu} G^A_{\mu\nu} \left(\frac{\sin\alpha}{\sin\beta}\right)^2 I_q\left[\frac{m_H^2}{4m_Q^2}\right]$};
\node[right,blue,xshift=5ex] (bt) at (b) {$y_{ij}\frac{m_{i,j}}{v}\bar L_i e_j$};
\draw[->,thick,blue,shorten >=2pt] (at) -- (a);
\draw[->,thick,blue,shorten >=2pt] (bt) -- (b);
\end{tikzpicture}
\caption{Heavy Higgs and heavy quarks integrated out
\label{fig:heavy-H}}
\end{subfigure}

\vspace{3ex}

\begin{subfigure}{\linewidth}\centering
	\begin{tikzpicture}[baseline=($(b)!0.5!(d)$)]
	\begin{feynman}
		\vertex (a);
		\vertex[below left=1cm of a]  (g1) {$g$};
		\vertex[above=of a] (c);
		\vertex[above left=1cm of c] (g2) {$g$};
		\vertex[right=of a] (b);
		\vertex[above=of b] (d);
		\vertex[below right=1cm of b] (f1) {$\ell^\prime$};
		\vertex[above right=1cm of d] (f2) {$\ell$};
		\diagram {
				
				(g1) -- [gluon] (a)  -- [fermion,edge label'=\(t\)] (c) -- [gluon] (g2),
				(a) --[fermion, edge label=\(t\)] (b) --[anti fermion] (f1),
				(f2) -- [fermion] (d)  -- [anti fermion,edge label=\(t\)] (c),
				(b) -- [charged scalar, edge label'=\(\frac{1}{m_X^2}\)] (d),
	};
\end{feynman}
\end{tikzpicture}
$\implies$
\begin{tikzpicture}[baseline=(b)]
		\begin{feynman}
			\vertex[blob] (b) {};
			\vertex[above left=of b] (a);
			\vertex[below left=of b] (c);
			\vertex[above left=1cm of a] (g1) {$g$};
			\vertex[below left=1cm of c] (g2) {$g$};
			\vertex[below right=of b] (f1) {\(\ell^\prime\)};
			\vertex[above right=of b] (f2) {\(\ell\)};
			\diagram* {
				
				(g1) -- [gluon] (a)  -- [fermion,edge label'=\(t\)] (c) -- [gluon] (g2),
				(a) --[fermion, edge label=\(t\)] (b) --[anti fermion] (f1) [particle=\(\ell^\prime\)],
				(f2) [particle=\(\ell\)] -- [fermion] (b)  -- [anti fermion,edge label=\(t\)] (c),
	};
\end{feynman}

\end{tikzpicture}

\caption{Heavy vector leptoquark and heavy quarks integrated out
\label{fig:heavy-LQ}}
\end{subfigure}
\caption{Examples of UV completions of dimension-8 operators with two gluon field strength tensors and two leptons.}
\label{fig:UV}
\end{figure}
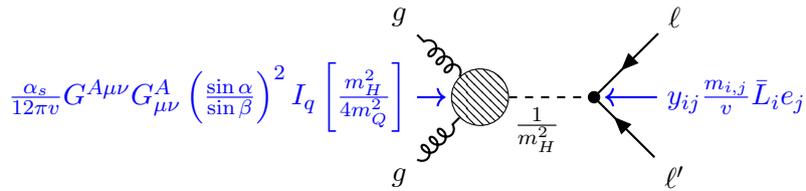
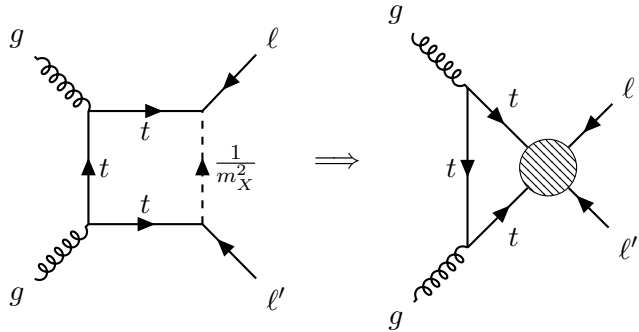

The spin zero operators are produced by a scalar
exchange in the s-channel as shown schematically
in Fig.~\ref{fig:heavy-H}. They can also be
produced  by a leptoquark exchange as
illustrated in Fig.~\ref{fig:heavy-LQ}
\cite{Potter:2012yv}. The simplest UV completion
scenario is a multi-Higgs model. The process in
Fig.~\ref{fig:heavy-H} would start from the
one-loop production of a heavy neutral Higgs from
gluon fusion which is dominated by the contributions of heavy quarks $Q$. This vertex is proportional to
known loop factors and mixing angles in the
scalar sector. The loop factor (indicated in the
figure as $I(m_H^2/(4m_Q^2)$) is known to be of
order one in the range $m_H \lesssim 2 m_Q$~\cite{Djouadi:2005gj}. The amplitude is not
suppressed by powers of $m_Q$  provided that the
$HQQ$ coupling is proportional to $m_Q$. An
example of a model with the necessary ingredients
is that of 
Refs.~\cite{BarShalom:2011zj,He:2011ti,Bar-Shalom:2016ehq}.
In this case the heavy neutral scalar gives mass
to the fourth generation quarks running in the
loop, we assume two of them for the factor of 2
in Eq.~\eqref{heavyH}. The second ingredient
necessary for this process to occur is a LFV coupling to the heavy scalar.
Such couplings are generic to multi-Higgs models
unless they are specifically removed with the use
of discrete symmetries. They have been studied
recently in the context of the $h\to \mu\tau$
limit set by CMS
\cite{Khachatryan:2015kon,CMS:2016qvi} in a
variety of multi-scalar models \cite{Crivellin:2015mga,Dorsner:2015mja,Altmannshofer:2015esa,Herrero-Garcia:2016uab,Hayreter:2016aex,Herrero-Garcia:2017xdu}. In the
notation of Ref.~\cite{Bar-Shalom:2016ehq}, the Wilson coefficients are 
generically given by 
\begin{equation}
	x_{ij} \sim  \frac{8\pi}{3}\left(\frac{1}{4\pi v}\right)^2 \frac{1}{m_H^2} \left(\frac{\cos\alpha\sin\beta}{\sin2\beta}\right)^2 \sin(\beta-\alpha)\Sigma^\ell_{ij} \frac{m_\ell}{v}.
	\label{heavyH}
	\end{equation}
The different factors involving $\alpha,\beta$
represent mixing in the scalar sector of the
model, $\Sigma^\ell_{ij}$ encodes the LFV Yukawa
coupling of the heavy scalar and $m_\ell$ is a
generic lepton mass associated with the latter.
The high energy scales suppressing this operator
can thus be as low as $\Lambda^4 \sim m_H^2 (4\pi
v)^2$ in this case.

The spin zero operators can also be produced by a
leptoquark as illustrated in
Fig.~\ref{fig:heavy-LQ}. The size of this
contribution can be estimated as follows: if the
leptoquark (scalar or vector) is very heavy,
part of the diagram can be contracted to a
four-fermion interaction as indicated in the
figure. This four-fermion interaction can then be
Fierz re-arranged to obtain the
$\bar{Q}Q\bar{\ell} \ell'$ structure. At this
point the diagram looks like the heavy scalar
exchange diagram and produces similar results
provided the same approximations hold. This
requires heavy fermions in the loop (heavier than
the top-quark) and the condition $\hat{s}
\lesssim 4m_Q^2$ for the loop factor to be of order
1.  The Fierz rearrangement also produces the
structure $\bar{Q}\gamma_5Q\bar{\ell}
\gamma_5\ell'$ which makes the diagram behave
like an exchange of a heavy pseudo-scalar and
produce the $x^\prime_{ij}$ coefficients. An
example of a model with the required ingredients
can be found in the discussion of heavy
vector-like quarks of
Ref.~\cite{Dobrescu:2016pda} which would yield 
\begin{equation}
x_{ij} \sim  x^\prime_{ij}\sim \frac{\pi}{3} \left(\frac{1}{4\pi v}\right)^2 \frac{1}{m_X^2} \lambda_\chi^2 \frac{v^2}{m_Q m_\ell}
	\label{leptoQ}
	\end{equation}
where $\lambda_\chi$ is the Yukawa interaction
between the  (scalar) leptoquark, the heavy
vector-like quark, and the lepton and it could be
of order one. The last factor rescales this
diagram to the one produced by a heavy Higgs.

Finally, the spin two operators can be produced by the tree-level exchange of a spin two resonance. One possibility is a composite analogue to the $f_2$ or $a_2$ resonances that occur in the strong interaction. A second possibility discussed in the literature is the exchange of Kaluza-Klein (KK) states \cite{Han:1998sg} where we can read directly from Eq.~(71) of this reference that, for example for $n=3$, 
\begin{eqnarray}
y_{ij}=z_{ij}=\frac{8\pi}{\alpha_s M_S^4} \delta_{ij} .
\label{KKexchange}
\end{eqnarray}
KK states would only produce parity and flavour conserving operators so they are not directly relevant for our study, but they illustrate how the operators may arise in a more complicated model.


\section{LFV signals at the LHC}\label{sec:LHC}
The operators listed in Eq.~\eqref{eq:ops} 
also produce distinctive signals with flavour violating (charged) lepton pairs at hadron colliders. We consider one operator at a time. Obviously in realistic UV completions multiple operators are generically produced at the same time, but the constraints in the "one-operator-at-a-time" approach also provide a good and simple estimate for the general strength of the constraints.

In the case where no spin correlations are observed, the spin averaged matrix elements squared produced by these operators are given by
\begin{align}
\overline{|\mathcal{M}_{X^{(\prime)}}|^2} &\sim \alpha_s^2 |x_{ij}^{(\prime)}|^2 v^2 \hat{s}^3 &
\overline{|\mathcal{M}_{{\bar X}^{(\prime)}}|^2} &\sim \alpha_s^2 |\bar x_{ij}^{(\prime)}|^2 v^2 \hat{s}^3 
\end{align}
for the operators without derivatives. The ratio of their cross sections are simply given by the ratio of the squared Wilson coefficients. The derivative operators on the other hand are more sensitive to the center of mass energy  and scale like
\begin{align}
\overline{|\mathcal{M}_Y|^2} &\sim \alpha_s^2 |y_{ij}|^2 \hat{s}^4 &
\overline{|\mathcal{M}_Z|^2} &\sim \alpha_s^2 |z_{ij}|^2 \hat{s}^4\;.
\end{align}
Again the ratio of the cross sections is given by $\sigma_Y/\sigma_Z = |y_{ij}|^2/|z_{ij}|^2$. This indicates that LHC limits cannot distinguish between the first four types of operators $\mathcal{O}_X^{(\prime)}$ and $\bar{\mathcal{O}}_X^{(\prime)}$ or between the last two, and that the LHC should be more sensitive to the latter ones. 
We quote all limits in terms of the scale $\Lambda$ of the operators which is given by the inverse of the fourth root of each Wilson coefficient, e.g. $\Lambda = x^{-1/4}$ for the operator $\mathcal{O}_X$.

For final states with $e\mu$ pairs, CMS has recently updated their analysis~\cite{CMS:2017usw} based on a data set of 35.9 fb$^{-1}$ at $\sqrt{s}=13$ TeV, 
which has improved the limits reported by ATLAS previously in Ref.~\cite{Aaboud:2016hmk}.
For $e\tau$ and $\mu\tau$ final states, however, the ATLAS search~\cite{Aaboud:2016hmk} still places the most stringent limit with 3.2 fb$^{-1}$ data 
at $\sqrt{s}=$13 TeV.    

The signal samples are generated at parton level with \texttt{MadGraph5\char`_aMC@NLO}~\cite{Alwall:2014hca}
and subsequently processed with \texttt{PYTHIA 8} for shower and hadronization~\cite{Sjostrand:2007gs}.
Note that no next-to-leading order factors are included\footnote{K-factors can be calculated using 
similar procedures as in  Ref.~\cite{Dawson:1990zj}} and the limits presented here are conservative. 
A fast detector simulation is performed using \texttt{Delphes}~\cite{deFavereau:2013fsa} with the default CMS configuration
for $e\mu$ final states and the ATLAS one for $e\tau$ and $\mu\tau$ final states.
\begin{figure}[tbp]
\centering
\includegraphics[width=0.45\textwidth]{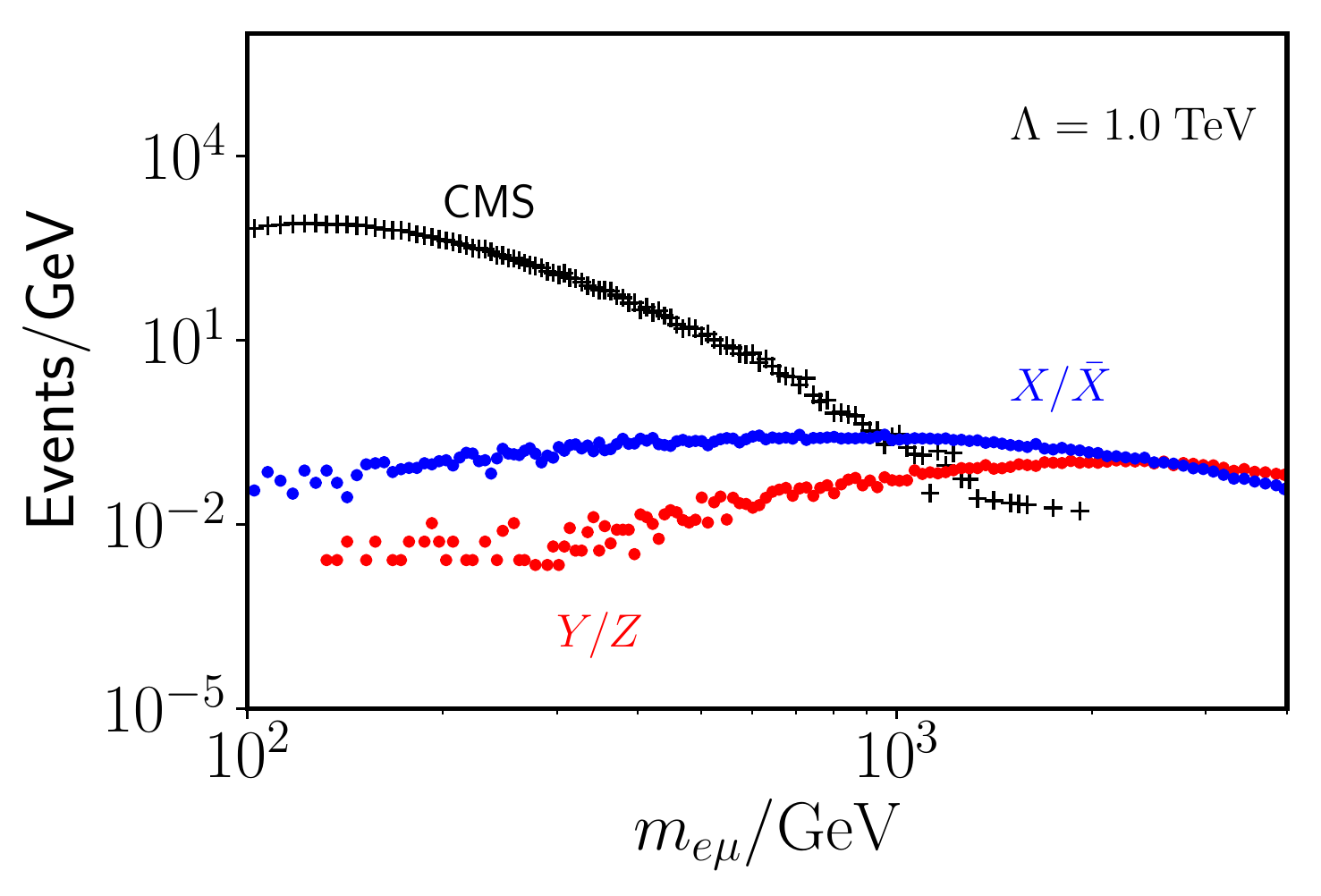}\hfill
\includegraphics[width=0.45\textwidth]{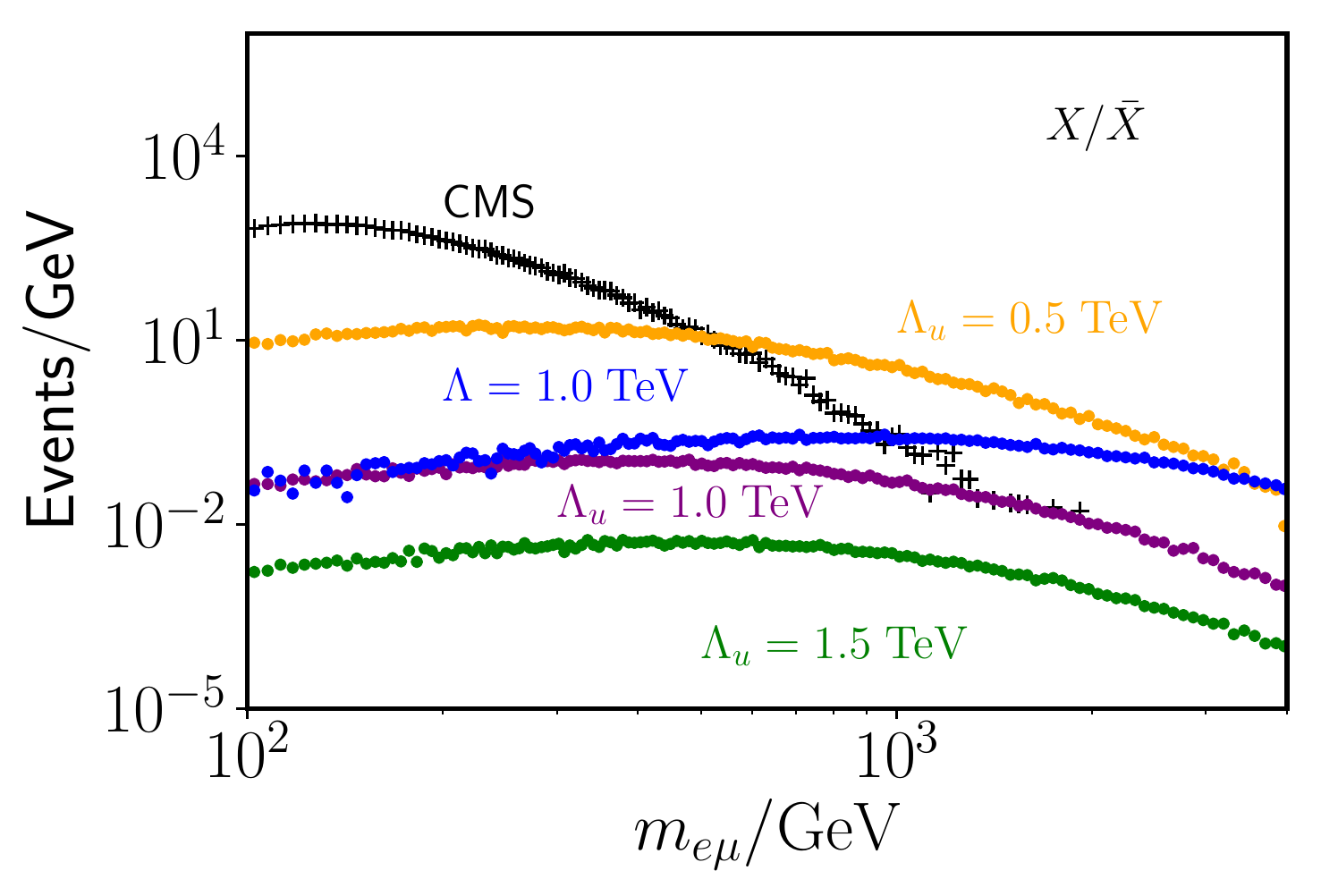}
\caption{The invariant mass distribution of selected $e\mu$ pairs for the signals together with CMS data shown in black.
With $\Lambda=1$ TeV and no unitarisation, signals in the left panel are in blue for $\mathcal{O}_{X}^{(\prime)}$, $\bar{\mathcal{O}}_{X}^{(\prime)}$ and in red for $\mathcal{O}_{Y,Z}$.
In the right panel, signals for operator $\mathcal{O}_X$ at $\Lambda=1$ TeV without unitarisation and $\Lambda_u=0.5$, $1$ and $1.5$ TeV with unitarisation are shown in orange, purple and green respectively. }
\label{fig:dist}
\end{figure}

In the left panel of Fig.~\ref{fig:dist} we show the predicted invariant mass distribution of $e\mu$ pairs at the reconstruction level 
together with the data from CMS in black for operators $\mathcal{O}_{X}^{(\prime)}$, $\bar{\mathcal{O}}_X^{(\prime)}$
and $\mathcal{O}_{Y,Z}$ in blue and red respectively with $\Lambda=1$ TeV in the simple\footnote{We use "simple" to clearly distinguish the EFT from the unitarised EFT. The scattering amplitude is not unitarised for the simple EFT.} EFT. Note that the cross section is strictly proportional to $\Lambda^{-8}$ in this case
and the shape of the distribution does not change. So it is straightforward to derive the lower limit on $\Lambda$ 
from the upper limit on the production cross section for simple EFTs. 
Clearly the distribution of the signal does not decline as fast as the SM background 
in the high energy region due the enhancement from the derivatives in the operators. 

As is well known the EFT calculation eventually violates perturbative unitarity at some energy and become an obvious over-prediction of the rate. This can be a problem in calculations for the LHC, where the parton distribution functions (PDF's) allow the process to probe energies as large as $\sqrt{s}$, albeit with decreasing probability. To address this issue, we follow the ad-hoc prescription described in  Ref.~\cite{Baur:1993fv} and replace the couplings of the effective operators by a form factor 
\begin{equation}
C \to \frac{C}{1+\tfrac{\hat s}{\Lambda^{\prime 2}}}
\end{equation}
where $\Lambda^\prime$ is an arbitrary scale which we choose to be equal to
the scale $\Lambda$ of the operator\footnote{$\Lambda^\prime$ does not have to be equal to the
scale $\Lambda$ of the operator in general.}. This ensures that the predicted
cross sections do not violate perturbative unitarity anywhere and leads to
meaningful limits on the effective operators. In the following we denote
the scale of the operator in the unitarised EFT by $\Lambda_u$ and the scale of
the operator in the simple EFT by $\Lambda$. 
In the right panel of Fig.~\ref{fig:dist}, we show the invariant mass distribution for operators $\mathcal{O}_X$ without 
unitarisation at $\Lambda=1$ TeV in blue and with unitarisation at $\Lambda_u = 0.5$, $1$ and $1.5$ TeV in orange, purple and green, respectively. 
\begin{figure}[tb]
\centering
\includegraphics[width=0.45\textwidth]{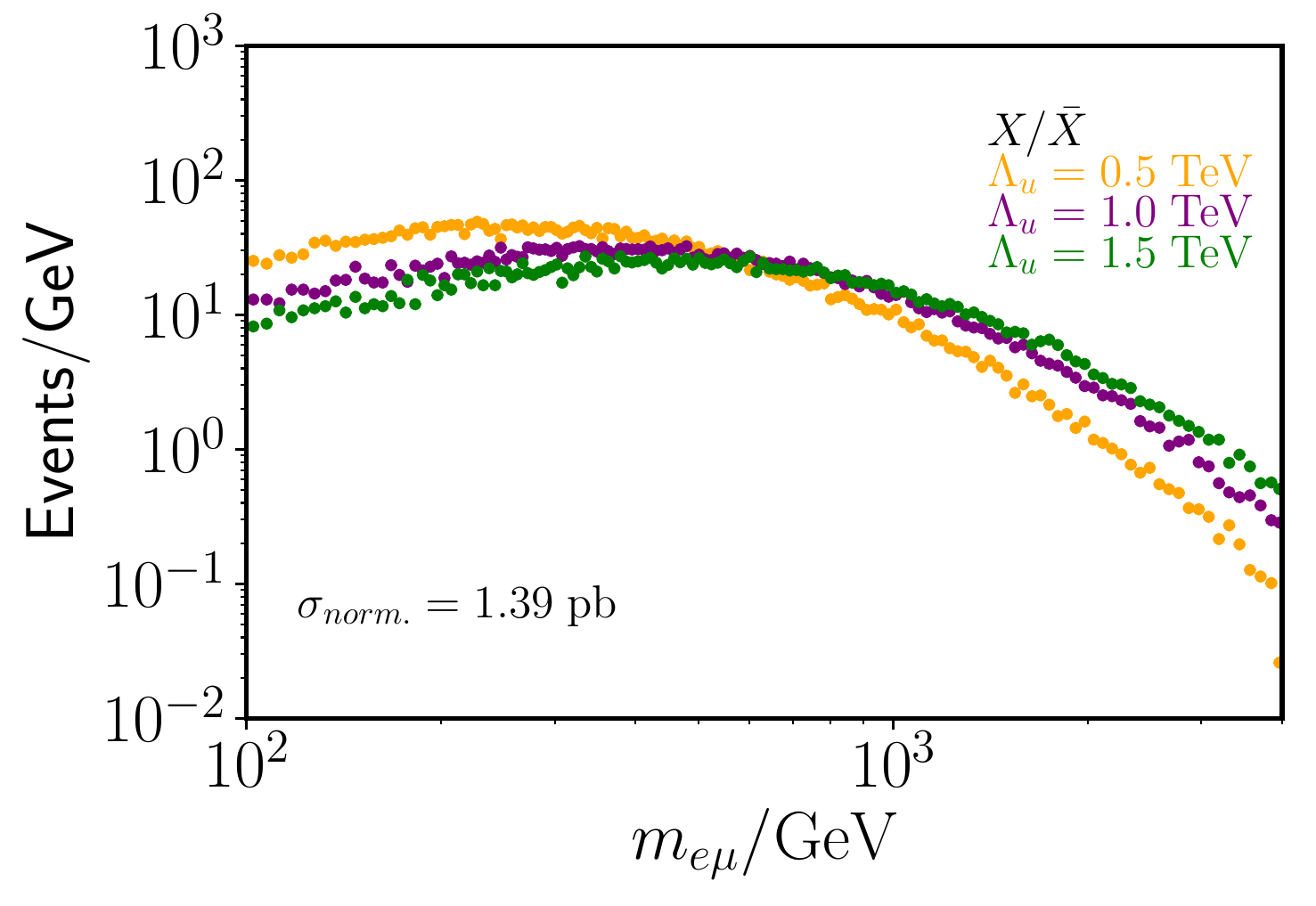}\hfill
\includegraphics[width=0.45\textwidth]{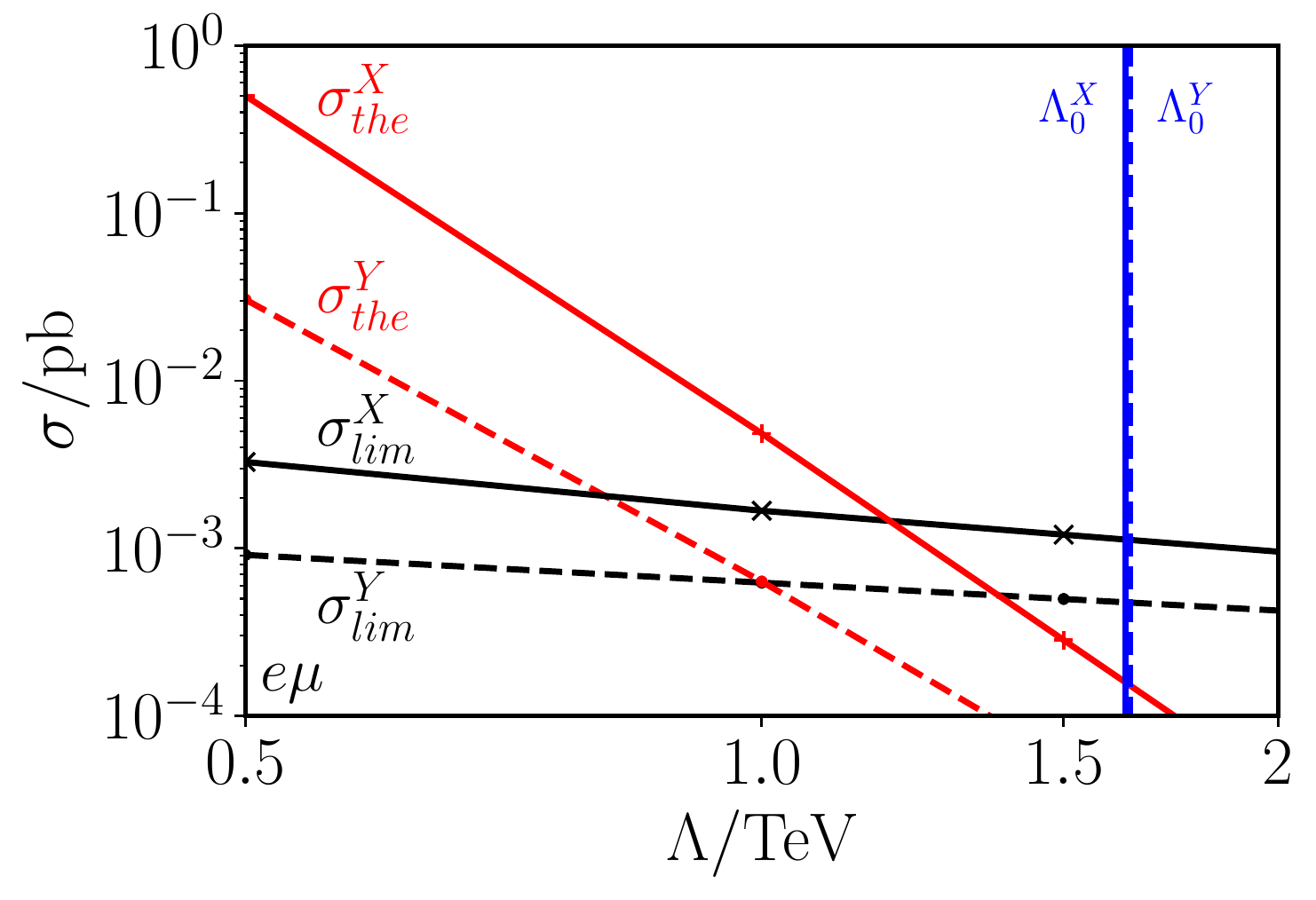}
\caption{Left: The invariant mass distribution of selected $e\mu$ pairs for $\mathcal{O}_{X}^{(\prime)}$, $\bar{\mathcal{O}}_X^{(\prime)}$ with $\Lambda_u=0.5$, $1$ and $1.5$ TeV normalised to the same cross section. Right: The blue solid (dashed) line shows the upper limit on the production cross section at $95\%$ C.L. for the operator $\mathcal{O}_X$ ($\mathcal{O}_Y$) in the simple EFT. The red solid (dashed) line denotes the predicted cross section for the operator $\mathcal{O}_X$ ($\mathcal{O}_Y$) in the unitarised EFT. The upper limit in the unitarised EFT is shown as a black solid (dashed) line for the operator $\mathcal{O}_X$ ($\mathcal{O}_Y$). The lower limit on $\Lambda_u$ is the intersection of the red and black lines.}
\label{fig:emulimit}
\end{figure}%
\begin{figure}[tb]
\centering
\includegraphics[width=0.45\textwidth]{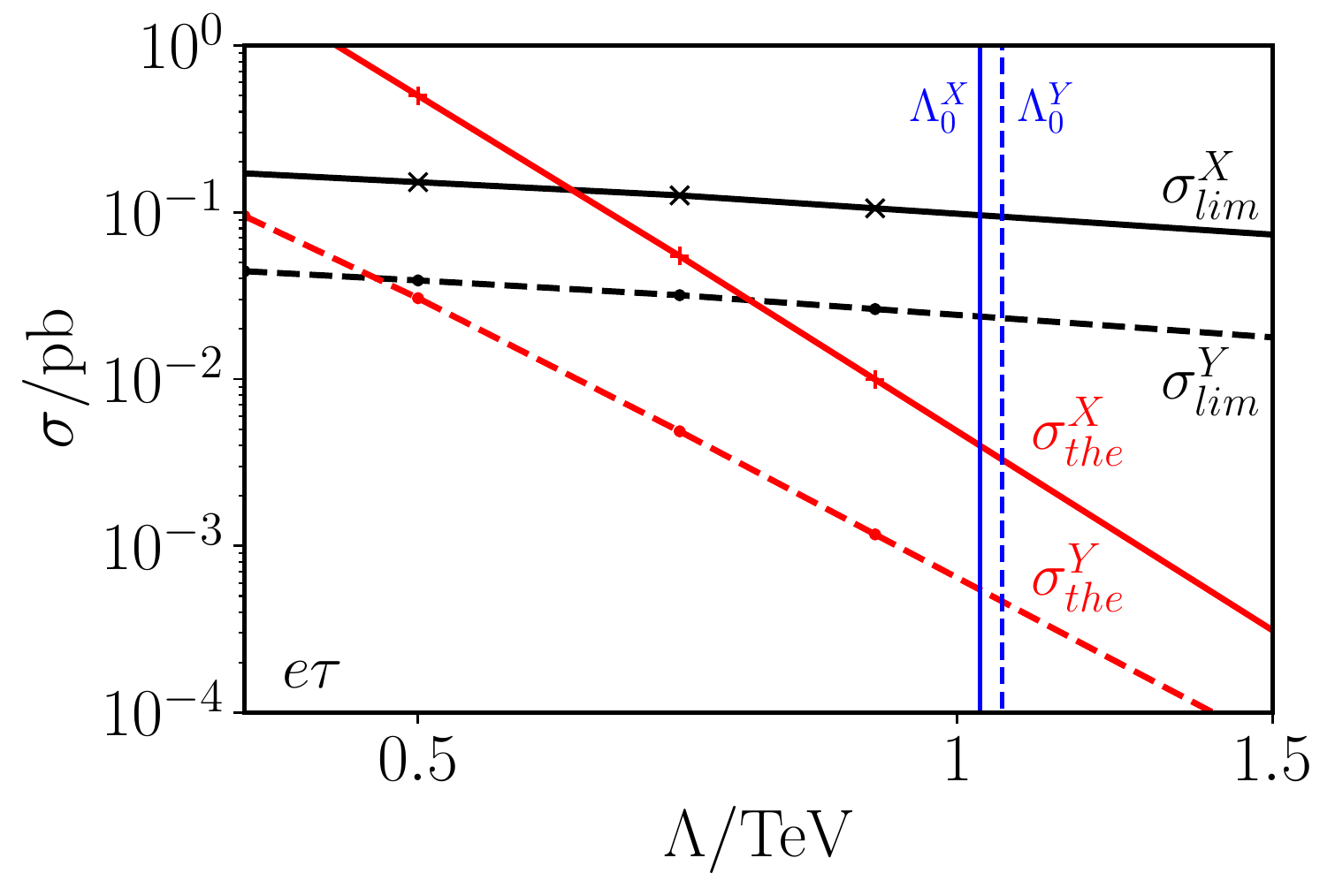}\hfill
\includegraphics[width=0.45\textwidth]{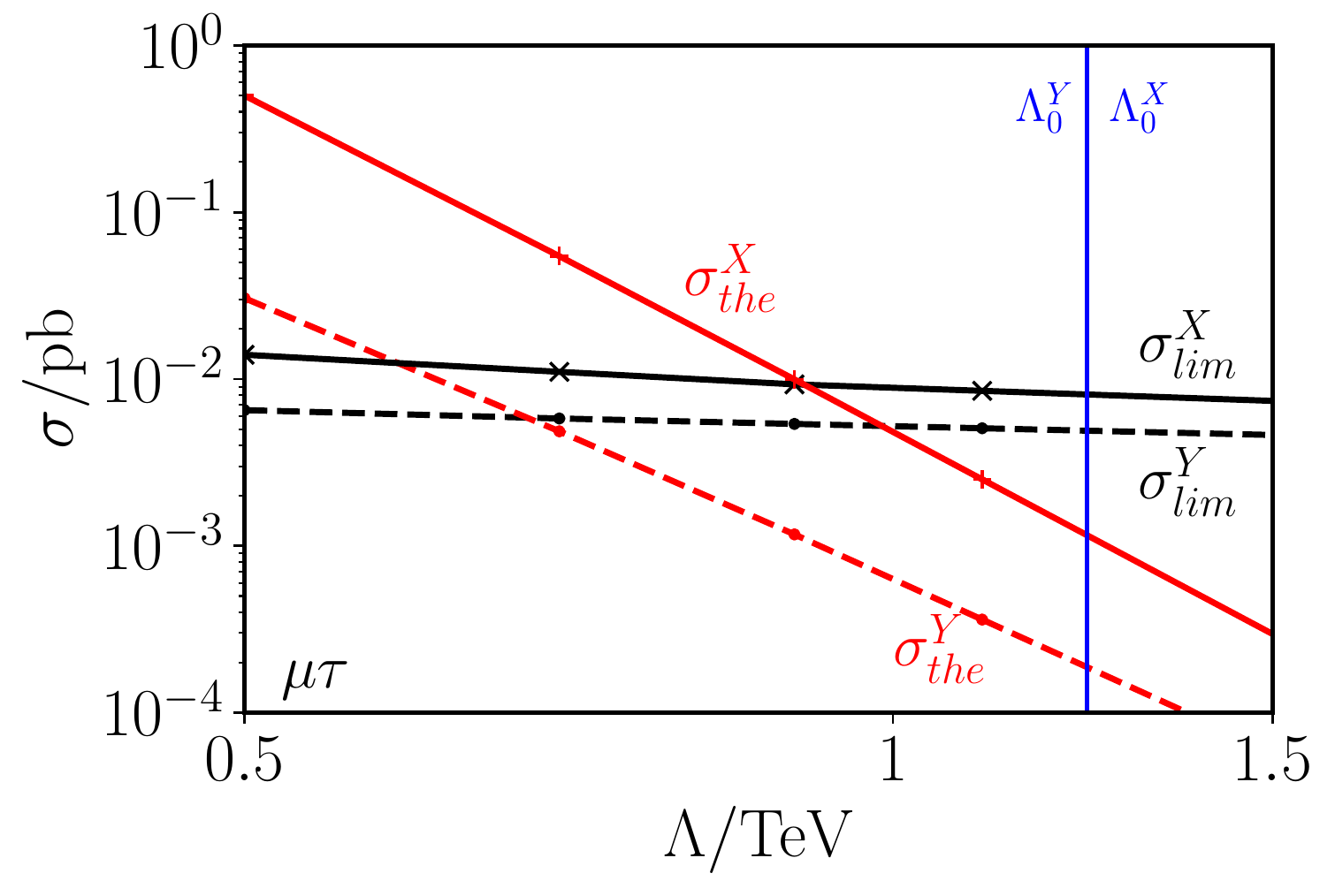}
\caption{ The colours of the lines are the same as in the right part of Fig.~\ref{fig:emulimit}.  Left: Operators with flavours $e\tau$. Right: Operators with flavours $\mu\tau$. The limits for the simple EFT lie on top of each other. Note the different ranges for $\Lambda$.}
\label{fig:atlas}
\end{figure}%
The CMS data is shown in  black again. Compared with the simple EFT without unitarisation, the high energy tail of the distribution is 
tamed as it should be. If we further increase $\Lambda_u$ to 1.5 TeV or larger we are increasing the scale of the operator 
affecting the result in two ways: the production cross section decreases; but there are relatively more high energy events 
as shown in Fig.~\ref{fig:emulimit}.
In this case the distribution is changing with $\Lambda_u$ and in order to derive the constraints 
we need to vary $\Lambda_u$ until the predicted cross section matches the upper limit derived from the distribution.    
\begin{table}[t]
\centering
\begin{tabular}{c|cc|cc}
\hline
& \multicolumn{2}{c|}{$\Lambda$ [TeV]} & \multicolumn{2}{c}{ $\Lambda_u$ [TeV]}\\
 & $\mathcal{O}_{X}^{(\prime)}$, $\bar{\mathcal{O}}_{X}^{(\prime)}$  & $\mathcal{O}_{Y,Z}$  & $\mathcal{O}_{X}^{(\prime)}$, $\bar{\mathcal{O}}_X^{(\prime)}$ & $\mathcal{O}_{Y,Z}$   \\
\hline\hline
$e\mu$   & 1.63 & 1.64 & 1.17 & 0.98 \\
$e\tau$   & 1.03 & 1.06 & 0.61  & 0.47  \\
$\mu\tau$ & 1.23 & 1.23 & 0.90 & 0.67  \\
\hline
\end{tabular}
\begin{minipage}{0.6\linewidth}\caption{Lower limits on the scale for various operators with ($\Lambda_u$) and without ($\Lambda$) unitarisation.}
\label{tab:LHCresults}
\end{minipage}
\end{table}

To derive the lower limit on $\Lambda$, we first construct the negative log likelihood function
assuming the number of events in each bin follows a Poisson distribution while the errors follow a Gaussian distribution. Both the data and the background distributions are taken from the experimental papers~\cite{CMS:2017usw,Aaboud:2016hmk}.
Then the profiled likelihoods are calculated taking the errors as nuisance parameters.
The upper limit on the production cross section at 95\% confidence level (C.L.) corresponding to a specific shape of distribution is achieved at $\Delta(-2 \ln L) = 3.84$ from the minimum of the negative log likelihood.
With the upper limits derived, the lower limit on $\Lambda$ in the simple EFT without unitarisation can be easily calculated.
For the unitarised case, we need to vary $\Lambda_u$ to find where the predicted production cross section just meets the upper limit on the cross section 
for the corresponding shape of distribution as shown in Fig.~\ref{fig:emulimit}.
We find CMS can exclude scales $\Lambda$ of the operators below 1.63 and 1.64 TeV for $\mathcal{O}_{X,\bar{X}}$ and $\mathcal{O}_{Y,Z}$ with $e\mu$ in simple EFT, while
with unitarisation the exclusion limits are slightly lowered to 1.17 and 0.98 TeV respectively. 
The same procedure can be applied to the ATLAS search with $e\tau$ and $\mu\tau$ final states as shown in Fig.~\ref{fig:atlas}.
We have summarised the limits on the scales of the operators for $e\mu$, $e\tau$ and $\mu\tau$ final states in Tab.~\ref{tab:LHCresults}.


\section{Low-energy precision constraints}\label{sec:LE}
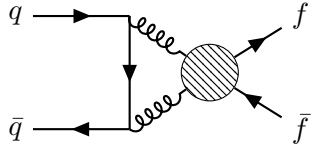
\begin{figure}[tb]\centering
	\begin{tikzpicture}
		\begin{feynman}
			\vertex (q1) {$q$};
		\vertex[below=of q1] (q2) {$\bar q$};
		\vertex[right=of q1] (g1) ;
		\vertex[right=of q2] (g2) ;
		\vertex[right=2 of g1] (f1) {$f$};
		\vertex[right=2 of g2] (f2) {$\bar f$};
		\vertex[blob,xshift=30]  (b) at ($(g1)!0.5!(g2)$) {};

\diagram*{
	(q1) -- [fermion] (g1) -- [fermion] (g2) -- [fermion] (q2),
	(g1) -- [gluon] (b) -- [fermion] (f1),
	(g2) -- [gluon] (b) -- [anti fermion] (f2),
	};
\end{feynman}
\end{tikzpicture}

	\caption{Operator mixing: $\bar q P_X q \bar\ell P_Y \ell$ induced at one-loop order}
	\label{fig:2q2l}
\end{figure}

In general the operators are also constrained by low-energy precision
measurements. The main constraints are from $\mu$-$e$ conversion in nuclei and from 
semi-leptonic $\tau$ decays. We focus on the operators without derivatives 
because operators with derivatives are further suppressed at low energies by a factor $E/\Lambda$, 
where $E$ is the typical energy of the process. This typical energy is the $\tau$ mass for $\tau$ decays or the
transferred momentum in $\mu$-$e$ conversion resulting in a suppression  of 
$E/\Lambda\lesssim 10^{-3}$.  The normalisation of the operators with an explicit $\alpha_s$
ensures that they are invariant under QCD corrections at one-loop order. This allows us 
to directly compare the LHC results with the low-energy constraints. Operators with two
quarks and two leptons are induced via the diagram in Fig.~\ref{fig:2q2l}, but are suppressed by a loop factor and by the quark mass in the loop, see e.g.~the discussion in
Ref.~\cite{Crivellin:2017rmk}. Thus we neglect
them in the following.
\begin{table}[tb]\centering
	\begin{tabular}{rlccc}
		\toprule
		nucleus & model  & $S^{(p)}$ [$m_\mu^{5/2}$] & $S^{(n)}$ [$m_\mu^{5/2}$] & $\Gamma_{\text{capt}}(\mu^- N)$ [$s^{-1}$] \\\midrule
		${}^{48}_{22}\mathrm{Ti}$ & FB & 0.0368 & 0.0435 & $2.59 \times 10^{6}$   \\
		${}^{197}_{\;\,79}\mathrm{Au}$ & 2pF & 0.0614 & 0.0918 & $13.07\times 10^{6}$   \\
		\bottomrule
\end{tabular}
\caption{Relevant assumptions for $\mu$-$e$ conversion in nuclei. $S^{(p)}$ and $S^{(n)}$ denote the overlap integrals~\cite{Kitano:2002mt} and $\Gamma_{\text{capt}}$ the capture rate of $\mu^-$ in the nucleus~\cite{Kitano:2002mt,Suzuki:1987jf}.}
	\label{tab:mueconv}
\end{table}

\begin{table}[tb]\centering
	\begin{tabular}{lc|lc}
	\toprule
	process & exp.~limit & operator & $\Lambda$ [TeV]\\
\midrule\midrule
	\multicolumn{4}{c}{$e \mu$}\\\midrule
	Br($\mu^-\, {}^{48}_{22}\mathrm{Ti} \to e^-\, {}^{48}_{22}\mathrm{Ti}$) & $<4.3\times 10^{-12}$ & $\mathcal{O}_X$, $\bar{\mathcal{O}}_X$  & $2.11$\\
	Br($\mu^-\, {}^{197}_{\;\,79}\mathrm{Au} \to e^-\, {}^{197}_{\;\,79}\mathrm{Au}$) & $<7\times 10^{-13}$ & $\mathcal{O}_X$, $\bar{\mathcal{O}}_X$  & $2.54$\\
		\midrule
	\multicolumn{4}{c}{$e \tau$}\\\midrule
	Br($\tau^+\to e^+ \pi^+ \pi^-$) & $<2.3\times 10^{-8}$ & $\mathcal{O}_X$, $\bar{\mathcal{O}}_X$   & $0.42$\\
	Br($\tau^-\to e^- K^+ K^-$) & $<3.4\times 10^{-8}$ & $\mathcal{O}_X$, $\bar{\mathcal{O}}_X$   & $0.37$\\
Br($\tau^-\to e^-\eta$) & $<9.2\times 10^{-8}$ & $\mathcal{O}_X^\prime$, $\bar{\mathcal{O}}_X^\prime$  & $0.40$\\
	Br($\tau^-\to e^-\eta^\prime$) & $<1.6\times 10^{-7}$ & $\mathcal{O}_X^\prime$, $\bar{\mathcal{O}}_X^\prime$  & $0.44$\\
	\midrule
	\multicolumn{4}{c}{$\mu\tau$}\\\midrule
	Br($\tau^-\to \mu^- \pi^+ \pi^-$) & $<2.1\times 10^{-8}$ & $\mathcal{O}_X$, $\bar{\mathcal{O}}_X$  & $0.43$\\
	Br($\tau^-\to \mu^- K^+ K^-$) & $<4.4\times 10^{-8}$& $\mathcal{O}_X$, $\bar{\mathcal{O}}_X$ & $0.36$\\
	Br($\tau^-\to \mu^-\eta$) & $<6.5\times 10^{-8}$ & $\mathcal{O}^\prime_X$, $\bar{\mathcal{O}}_X^\prime$& $0.42$\\
	Br($\tau^-\to \mu^-\eta^\prime$) & $<1.3\times 10^{-7}$  & $\mathcal{O}^\prime_X$, $\bar{\mathcal{O}}_X^\prime$& $0.46$ \\
\bottomrule
\end{tabular}

\caption{Constraints on the couplings from low-energy precision experiments. The limits are taken from Ref.~\cite{Patrignani:2016xqp}. }
\label{tab:LEconstraints}
\end{table}
As discussed in Ref.~\cite{Petrov:2013vka} coherent $\mu$-$e$ conversion in nuclei and semi-leptonic $\tau$ decays constrain different combinations of the Wilson coefficients of the form
\begin{equation}
	|\mathcal{X}^{(\prime)}_{ij}|^2\equiv  |x^{(\prime)}_{ij}|^2 + |\bar x^{(\prime)}_{ij}|^2 + |x^{(\prime)}_{ji}|^2 + |\bar x^{(\prime)}_{ji}|^2\;.
\end{equation}
The branching ratio of the $\mu$-$e$ conversion rate in a nucleus $N$ over the capture rate of a muon $\Gamma_{\rm capt} (\mu^- N)$ is~\cite{Kitano:2002mt,Petrov:2013vka}
\begin{equation}
	\mathrm{Br}(\mu^- N \to e^- N) = \frac{64 \sqrt{2}\pi^2 }{81\, G_F\, \Gamma_{\rm capt} (\mu^- N)} \left| G_p S^{(p)} + G_n S^{(n)}\right|^2 |\mathcal{X}_{e\mu}|^2  
\end{equation}
with the Fermi constant $G_F=1/(\sqrt{2} v^2)$. The nuclear matrix element for a nucleon $\mathcal{N}=p,n$ is given by~\cite{Cheng:2012qr}
\begin{equation}
	G_\mathcal{N}\equiv \braket{\mathcal{N}|\frac{\alpha_s}{4\pi} G_{\mu\nu}^a G^{\mu\nu a} |\mathcal{N}} =-189\, \mathrm{MeV}
\end{equation}
using the strange quark sigma term $\sigma_s=m_s\braket{p|\bar s s |p}=50$ MeV. $S^{(p)}$ and $S^{(n)}$ are the overlap integrals, which are given in Tab.~\ref{tab:mueconv} together with other relevant parameters. We do not consider the parity-violating operators in Eq.~\eqref{eq:xbar} for $\mu$-$e$ conversion, because they induce spin-dependent $\mu$-$e$ conversion in nuclei, for which there are currently no strong constraints for two main reasons. 
First, the main light isotopes such as
${}^{48}_{22}$Ti do not have a nuclear spin; second the suppression of
spin-dependent $\mu$-$e$ conversion compared to spin-independent conversion is
stronger for heavier nuclei such as Au and Pb due to the absence of
coherent enhancement proportional to the number of nucleons squared. 
This is also justified by the final result: as the limit from coherent $\mu$-$e$ conversion in nuclei is of the same order as the constraint from the LHC, we do not expect that $\mu$-$e$ conversion can currently impose any competitive constraints for the derivative and parity-violating operators. However
spin-dependent $\mu$-$e$ conversion may become
an interesting probe in the future as it has been pointed out
in Refs.~\cite{Cirigliano:2017azj,Davidson:2017nrp}. The future COMET~\cite{Kuno:2013mha} and
Mu2e~\cite{Carey:2008zz} experiments may probe a
relevant region of parameter space for the parity-violating dimension-8 gluon operators which induce a spin-dependent pseudo-scalar
coupling to nucleons.

Operators with $\tau$ flavour are constrained by semi-leptonic $\tau$ decays. 
In particular the operators $\mathcal{O}_X$ and $\bar{\mathcal{O}}_X$ are constrained by parity-conserving $\tau$ decays to two light charged mesons.
In the limit of vanishing final state lepton mass the differential decay rate takes the compact form
\begin{equation}
	\frac{d\Gamma(\tau^-\to \ell^- M^+ M^-)}{dq^2} = \frac{m_\tau}{648\sqrt{2}\pi\, G_F} |\mathcal{X}_{\tau\ell}|^2 q^4 \sqrt{1-\frac{4m_M^2}{q^2}} \left(1-\frac{q^2}{m_\tau^2}\right)^2
\end{equation}
in terms of the momentum transfer to the meson system $q^2=(p_1+p_2)^2$.
The relevant hadronic matrix element for the parity-conserving operators is given by~\cite{Voloshin:1985tc,Shifman:1988zk}
\begin{align}
	\braket{M^+(p_1)M^-(p_2)|-\frac{9\alpha_s}{8\pi}G^{\mu\nu a} \tilde G^a_{\mu\nu}|0} & = (p_1+p_2)^2 \;. 
\end{align}
Numerical integration over $q^2$ from $4 m_M^2$ to $m_\tau^2$ yields the decay rate.

The parity-violating operators $\mathcal{O}_X^\prime$ and $\bar{\mathcal{O}}_X^\prime$ are constrained by semi-leptonic $\tau$ decays to a charged lepton and a neutral pseudo-scalar $M$.
The partial width of $\tau^-\to \ell^- M$ is given by
\begin{equation}
	\Gamma(\tau^-\to \ell^- M)	= \frac{\pi m_\tau}{\sqrt{2}\,G_F} |a_M|^2 |\mathcal{X}^\prime_{\tau\ell}|^2 \left(1-\frac{m_M^2}{m_\tau^2}\right)^2\;.
\end{equation}
with the hadronic matrix element 
\begin{align}
	\braket{M(p)|\frac{\alpha_s}{4\pi}G^{\mu\nu a} \tilde G^a_{\mu\nu}|0} & \equiv a_M\;.
\end{align}
For the $\eta$ and $\eta^\prime$ mesons the matrix elements $a_M$ are calculated in the Feldmann-Kroll-Stech (FKS) scheme~\cite{Feldmann:1998vh,Feldmann:1999uf} and read~\cite{Beneke:2002jn}
\begin{align}
a_{\eta} & = -\frac{m_{\eta^\prime}^2-m_\eta^2}{2}\sin2\phi \,\left(- \frac{f_q\sin\phi}{\sqrt{2}} +f_s \cos\phi\right) \approx-0.022\;,\\\nonumber 
a_{\eta^\prime} & = -\frac{m_{\eta^\prime}^2-m_\eta^2}{2}\sin2\phi \,\left(\frac{f_q\cos\phi}{\sqrt{2}} +f_s \sin\phi\right) \approx -0.056\;.
\end{align}
The three parameters in the FKS scheme are determined from a fit to experimental data~\cite{Feldmann:1998vh,Feldmann:1999uf}
\begin{align}
	f_q& = 1.07\, f_\pi\,, &
	f_s & = 1.34\, f_\pi\,, & 
	\phi & = 39.3^\circ 
\end{align}
in terms of the pion decay constant $f_\pi  = 130.2\, \mathrm{MeV}$~\cite{Patrignani:2016xqp}.

We summarise the constraints from the discussed processes in Tab.~\ref{tab:LEconstraints}. The table is
split into three parts, one for each combination of flavours. The first column
shows the relevant process and the current constraint for each
branching ratio~\cite{Patrignani:2016xqp} is given in the second
column. The third column indicates the operators which are constrained by
the process and the lower limit on the scale $\Lambda$ is given in the
fourth column.
The most stringent constraints are for lepton
flavours $e\mu$ from $\mu$-$e$ conversion in
nuclei: the scale of the operator $\Lambda$ has
to be larger than about $2.54$ TeV. Constraints on
$\tau$ flavour are about
one order of magnitude less stringent and lead to lower limits of order $400$ GeV.
The final limit on the scale $\Lambda$ of the operator is not very sensitive to the detailed nuclear and hadronic physics because the scale is the fourth root of the Wilson coefficient, $\Lambda = x_{ij}^{-1/4}$.


\section{Summary}
\label{sec:summary}
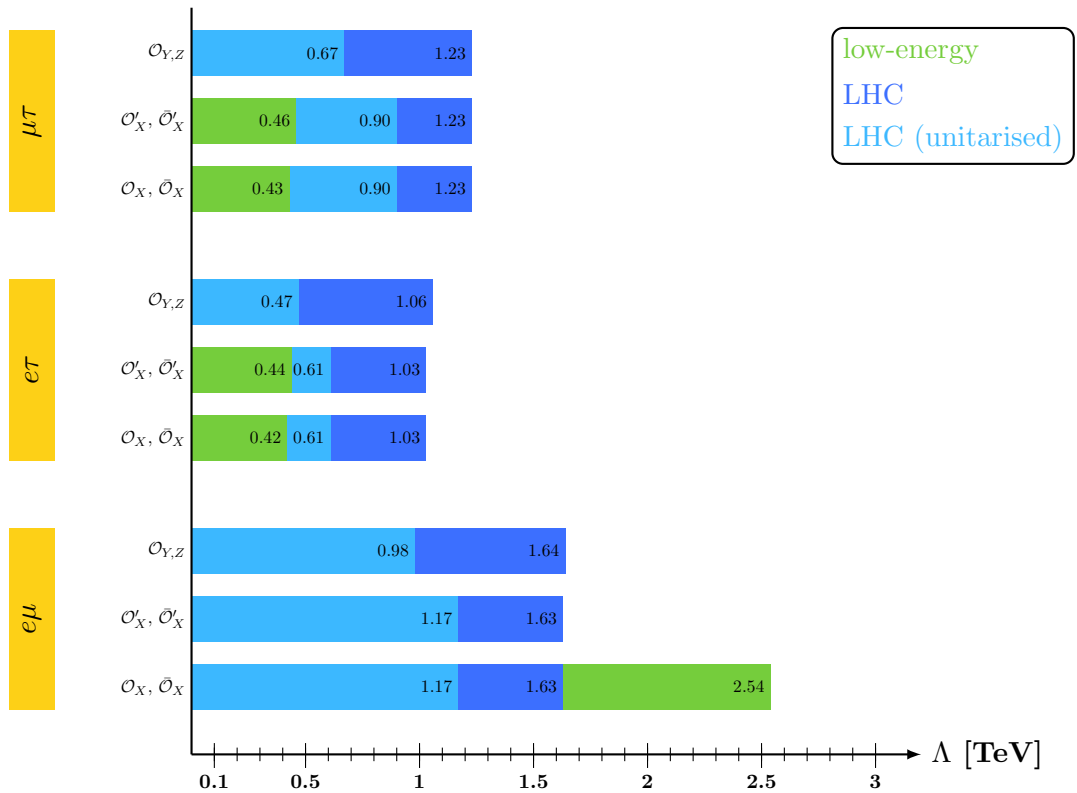
\begin{figure}[pb!]\centering
	\begin{tikzpicture}[x=3cm,y=3cm]
		\edef\mypos{-0.20}
		\barlimc{$\mathcal{O}_X$, $\bar{\mathcal{O}}_X$}{LE}{2.54}{LHC}{1.63}{LHCunit}{1.17}
	        \fill[legend] ($(-0.6,\mypos)+(0,-0.1)$) rectangle +(-0.2,0.8);
		\barlimb{$\mathcal{O}_X^\prime$, $\bar{\mathcal{O}}_X^\prime$}{LHC}{1.63}{LHCunit}{1.17}
		\node[rotate=90] at (-0.7,\mypos) {$e\mu$};
		\barlimb{$\mathcal{O}_{Y,Z}$}{LHC}{1.64}{LHCunit}{0.98}
		\edef\mypos{0.9}
			\barlimc{$\mathcal{O}_X$, $\bar{\mathcal{O}}_X$}{LHC}{1.03}{LHCunit}{0.61}{LE}{0.42}
		\fill[legend] ($(-0.6,\mypos)+(0,-0.1)$) rectangle +(-0.2,0.8);
			\barlimc{$\mathcal{O}_X^\prime$, $\bar{\mathcal{O}}_X^\prime$}{LHC}{1.03}{LHCunit}{0.61}{LE}{0.44}
		\node[rotate=90] at (-0.7,\mypos) {$e\tau$};
		\barlimb{$\mathcal{O}_{Y,Z}$}{LHC}{1.06}{LHCunit}{0.47}
		\edef\mypos{2.0}
		\barlimc{$\mathcal{O}_X$, $\bar{\mathcal{O}}_X$}{LHC}{1.23}{LHCunit}{0.90}{LE}{0.43}
		\fill[legend] ($(-0.6,\mypos)+(0,-0.1)$) rectangle +(-0.2,0.8);
		\barlimc{$\mathcal{O}_X^\prime$, $\bar{\mathcal{O}}_X^\prime$}{LHC}{1.23}{LHCunit}{0.90}{LE}{0.46}
		\node[rotate=90] at (-0.7,\mypos) {$\mu\tau$};
		\barlimb{$\mathcal{O}_{Y,Z}$}{LHC}{1.23}{LHCunit}{0.67}

		\node[below left,LE,yshift=2ex] (LE) at (3.5,\mypos) {low-energy};
		\node[below right,LHC] (LHC) at (LE.south west) {LHC};
		\node[below right,LHCunit] (LHCunit) at (LHC.south west) {LHC (unitarised)};
		\draw[rounded corners,draw=black,thick] (LE.north west) rectangle (LHCunit.south east);

		\draw[thick] (0,-0.2) -- +(0,3.3);
		\draw[thick,-latex] (0,-0.2) -- +(3.2,0);
		\node[right] at (3.2,-0.2) {\bf $\Lambda$ [TeV]}; 
	\foreach \x in {1,...,30}
	{        
		\draw ($(\x/10,-0.23)$) -- ($(\x/10,-0.17)$);
    }
		 \foreach \x in {0.1,0.5,1,1.5,2,2.5,3}
    {        
	    \draw (\x,-0.26) -- (\x,-0.15);
	    \node[below,scale=0.7] at (\x,-0.26) {\bf\x};
    }

	\end{tikzpicture}
	\caption{Summary of all constraints on the scale of the dimension-8 operators with two gluon field strength tensors and two leptons. Green indicates the most stringent low-energy constraint and limits obtained from the LHC study are shown in blue: dark blue for ordinary EFT and light blue for the unitarised EFT. The exact numerical value for each constraint is given at the end of each bar in TeV.}
	\label{fig:summary}
\end{figure}

We considered the six lepton-flavour-violating
gluonic dimension-8 operators which can be
generated in many models of physics beyond the
SM. As they induce new clean
processes at hadron colliders we used results
from the ATLAS and CMS experiments, the two general purpose
experiments at the LHC,
to obtain constraints on their effective scale
which is the main result of our study. As the LHC
energy $\sqrt{s}=13$ TeV is larger than the
obtained lower limits on the scales in the EFT and thus there may be a violation of perturbative unitarity we also
interpret the analysis in terms of a unitarised
EFT which provides a smooth cutoff. The limits obtained in the
unitarised EFT are lower, but of the same order of magnitude as the ones of the EFT.
The cross sections $\sigma$ scale with the EFT scale $\Lambda$ of the effective operator to the eighth power: $\sigma\propto\Lambda^{-8}$. Future studies at the LHC have the potential to improve the limits on $\Lambda$, but are limited by the strong dependence of the cross section on the EFT scale $\Lambda$. The constraint on $\Lambda_u$ in the unitarised EFT moves closer to the one on the EFT scale $\Lambda$ with increasing $\Lambda$, because the effect of the unitarisation is reduced for higher EFT scales.
Low energy precision experiments such as
$\mu$-$e$ conversion in nuclei and semi-leptonic
$\tau$ decays also provides constraints on the
scale of the operator. 

Fig.~\ref{fig:summary} summarises all
results. The LHC limits are shown in (light) blue for the (unitarised) EFT and the most stringent limit obtained from low-energy precision experiments is shown in green. The figure clearly demonstrates the
complementarity between constraints provided by
the LHC and low-energy precision experiments. The LHC generally provides the
most stringent constraint for all operators apart from parity-conserving operators of the form $GG\bar \mu P_{L,R} e$.
For lepton-flavour-violating gluonic
operators with $\tau$ leptons the LHC clearly
provides the most stringent limits.
For $e\mu$ flavour the constraints from the LHC are outperformed by the experimental limit from $\mu$-$e$ conversion in
${}^{197}_{79}\mathrm{Au}$ for the operators
$\mathcal{O}_X$ and $\bar{\mathcal{O}}_X$. For the other operators of $e\mu$ flavour
$\mu$-$e$ conversion in nuclei is suppressed. However the future 
$\mu$-$e$ conversion experiments 
COMET~\cite{Kuno:2013mha} and
Mu2e~\cite{Carey:2008zz} will dramatically improve the sensitivity to $\mu$-$e$ conversion by several orders of magnitude  and thus provide an interesting probe for
these operators as well.

\section*{Acknowledgements}
We thank Bogdan Dobrescu, Tao Han and Amarjit Soni for discussions of
their respective models. This work was supported in part by the
Australian Research Council.
All Feynman diagrams were generated using the Ti\textit{k}Z-Feynman
package for \LaTeX~\cite{Ellis:2016jkw}.

\section*{Note added}
While this work was in its final stages, a related
work~\cite{Bhattacharya:2018ryy} on testing LFV gluonic dimension-8
operators at the LHC was posted on the arxiv. Our studies are
complementary as Ref.~\cite{Bhattacharya:2018ryy} focuses on future
prospects at the LHC with $100\,\mathrm{fb}^{-1}$ data, while we obtain actual limits by recasting existing LHC analyses with $35.9\, \mathrm{fb}^{-1}$ ($e\mu$)~\cite{CMS:2017usw} and $3.2\, \mathrm{fb}^{-1}$ ($e\tau,\,\mu\tau$)~\cite{Aaboud:2016hmk} data. In addition we studied operators with derivatives and compared the current LHC limits to constraints from low-energy precision experiments. 

\small
\setlength{\bibsep}{0pt}
\bibliography{ref}

\end{document}